\documentclass{svproc}
\usepackage{xcolor}
\usepackage{amsmath,amssymb}
\usepackage{epsfig}
\usepackage{amssymb}
\usepackage{graphicx,float}
\usepackage{subcaption}
\usepackage{url}

\date{}
\begin{document}

\mainmatter

\title{Enhancing Anti-Money Laundering Efforts with Network-Based Algorithms}

\author{Anthony Bonato\inst{1}\thanks{Supported by an NSERC Discovery Grant.} \and Juan Sebastian Chavez Palan\inst{1,2} \and Adam Szava\inst{1}}

\institute{Toronto Metropolitan University \\
\and CIBC Capital Markets}

\maketitle

\begin{abstract}
The global banking system has faced increasing challenges in combating money laundering, necessitating advanced methods for detecting suspicious transactions. Anti-money laundering (or AML) approaches have often relied on predefined thresholds and machine learning algorithms using flagged transaction data, which are limited by the availability and accuracy of existing datasets. In this paper, we introduce a novel algorithm that leverages network analysis to detect potential money laundering activities within large-scale transaction data.  Utilizing an anonymized transactional dataset from Coöperatieve Rabobank U.A., our method combines community detection via the Louvain algorithm and small cycle detection to identify suspicious transaction patterns below the regulatory reporting thresholds. Our approach successfully identifies cycles of transactions that may indicate layering steps in money laundering, providing a valuable tool for financial institutions to enhance their AML efforts. The results suggest the efficacy of our algorithm in pinpointing potentially illicit activities that evade current detection methods.
\end{abstract}

\section{Introduction}\label{intro}

Financial regulation around the prevention of money laundering has evolved significantly over the past 50 years, due to both the globalization of the banking system and the vulnerability of the Internet to organized criminal activity. The first effort at enforcement came with the Bank Secrecy Act, signed in 1970 by the US Congress to prevent tax evasion through banks; see \cite{white}. Since then, many developed countries have created standalone governmental agencies to oversee their \emph{anti-money laundering} (or AML) mandates, such as the Financial Transactions and Reports Analysis Centre of Canada. There are also intergovernmental institutions, like the Financial Action Task Force (FATF), founded in 1989. The FATF is a coalition of 37 countries that align their policies to combat money laundering, corruption, tax evasion, and terrorist financing; see \cite{murrar}. These regulatory institutions rely on self-reported data provided by banks to enforce the inspection of compliance frameworks, risk management policies, and transaction monitoring of high-risk or illicit activity; see \cite{nojeim,schneider}. Banks, therefore, need to employ research and tools to help them identify which transactions they should report or face being sanctioned by these regulators.  

Studies on the identification of potential money laundering activity use various types of risk categorization models by employing algorithms that aim to differentiate between \emph{suspicious} and \emph{non-suspicious} transactions; see \cite{demetris}. The primary approach to anti-money laundering models involves analyzing flagged money laundering data through supervised machine learning algorithms; see \cite{yang}. There are limitations to these models, the most obvious being the prerequisite of having an initial set of flagged data, which is heavily reliant on the accuracy of existing money-laundering data, often private and therefore not abundantly available; see \cite{jullum}. Another challenge is that the amount of raw data obtained from financial institutions is extremely large and requires data refinement processes; this is why most of the studied methods use real datasets comprised of regular transactions only, while the flagged money laundering activities are artificially introduced; see \cite{chen}. Therefore, recent studies focus on unsupervised and semi-supervised machine learning algorithms. These are either based on categorizing suspicious transactions using fixed rules and thresholds provided by financial regulations or on unsupervised learning methods focused on cluster analysis; see \cite{yang}. 

We may view bank accounts and their transactions as a network, allowing us to study them using network science tools such as centrality measures and community clustering. Our objective in this paper is to analyze a real-world dataset collected from Coöperatieve Rabobank U.A.\ (or Rabobank), a Dutch multinational bank, and to describe a network-based algorithm that can assist with the detection of potential money laundering transactions without using artificially flagged accounts. The algorithm uses community and cycle detection to detect suspicious activity. A key novelty of our algorithm is its ability to identify a set of transactions currently not reported under existing threshold-based regulations, which can be used in conjunction with existing machine learning algorithms.

The paper is organized as follows. In the next section, we provide an overview of the AML model used in the banking industry. In Section~3, we describe the Rabobank networked dataset, followed by a discussion of the Louvain algorithm for community detection in Section~4. In Section~5, we present our AML algorithm leveraging network analysis, and we present our results in Section~6. We conclude with a discussion of future work.

We consider directed graphs (or \emph{digraphs}) with multiple directed edges in the paper. Additional background on graph theory may be found in \cite{west}, and more background on complex networks may be found in the book \cite{bonato}.

\section{The AML Model}\label{aml}

The money laundering model has not evolved much since the Bank Secrecy Act. Even though law enforcement and regulators look at different aspects, such as the origin of the money and the intent of its use, the industry still defines policies around the 1980s placement, layering, and integration model; see \cite{cassella}. This model describes a method for concealing cash behind legitimate business activities. To launder money, the owner of the illicit profit will try to infiltrate the legal banking network through these three steps; see \cite{schneider}.
\begin{enumerate}
    \item \emph{Placement}: After obtaining illegal profit, the first step is to find a way for the owner to deposit the cash into the financial system, perhaps through multiple deposits or by commingling it with the proceeds of a legitimate, cash-intensive business, such as a restaurant or gas station.
    \item \emph{Layering}: Once the cash is inside the financial system, the second step involves moving the money through a series of transactions to make the trail difficult to follow.
    \item \emph{Integration}: The final step involves using the money in a legitimate transaction to pay for goods and services, either to support the lifestyle of the original owner or to sustain the operation.
\end{enumerate}

With regard to combating placement, banks are required to maintain \emph{Know Your Client} (or KYC) profiles on each of their customers to understand the source of their income and determine the intended use of the bank account; see \cite{nojeim}. Layering is the hardest step to combat, as banks currently monitor transactions from accounts flagged as \emph{high-risk}, so most supervised AML algorithms rely on existing flagged accounts. High-risk accounts are those being owned or controlled by individuals or corporations that have been linked to any of the following: economic sanctions, political exposure, cash-intensive businesses, regulatory or criminal enforcement, or money laundering scandals; see \cite{schneider,white}. Another strategy to combat layering relies on threshold reporting. Currently, most countries have a defined amount for all banks to report a transaction to regulators; in the US and Canada, this amount is any transaction over \$10,000 (in the respective currency of the country); see \cite{siacca}.

The algorithm we will introduce in Section~5 focuses on the layering and integration steps of money laundering within a network of accounts without having prior knowledge of any illicit placement or flagged high-risk accounts. The intent is that the algorithm will be capable of searching across a large dataset of millions of bank account transactions to flag a potentially suspicious set of transactions that have moved less than the \$10,000 reporting threshold across different accounts, only to return a similar amount of money to the original account owner. It is worth noting that this does not confirm money laundering but identifies a set of potentially suspicious accounts that must be further investigated. Our algorithm can be used for an initial identification step, which can then be combined with existing AML machine-learning algorithms. The added benefit is that our algorithm will flag a set of initial suspicious accounts within real data instead of artificially introducing them. 

\section{Rabobank Data}

Rabobank was initiated by approximately 1,200 local cooperative banks, where people organized local banking administrations run by volunteers. The cooperatives’ governance structure, non-commercial ideology, and volunteer-based governance make them unique compared to other private banks in the country. In 1898, the Coöperatieve Centrale Raiffeisenbank and the Coöperatieve Centrale Boerenleenbank were founded to provide central organization and supervision. In 1972, they merged to form Rabobank; see \cite {degraaf}. Saxena et al.\ analyzed a digraph comprised of 1.6 million nodes with directed edges corresponding to transactions between the bank's users from 2010 to 2020; see \cite{saxena}.

\begin{figure}[ht!]
\centering
\includegraphics[width= 8 cm]{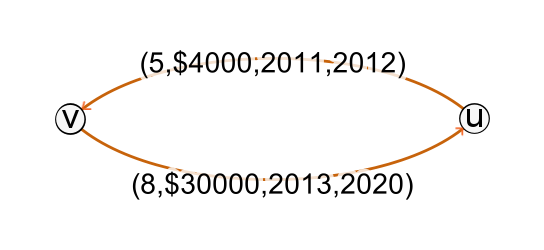}
\caption{Rabobank nodes and directed edges: two data elements $\mathbf{\vec{p}}_{1}(u,v)$ and $\mathbf{\vec{p}}_{2}(v,u)$.}\label{fig:f10}
\end{figure}

We used an anonymized Rabobank banking transaction dataset provided in \cite{saxena}. This dataset consists of bank accounts and transactions between them in an undisclosed currency. For simplicity, we list these as units of currency by dollars or \$. Each data element $\mathbf{\vec{p}}(u,v;n,k,Y_{1},Y_{2})$ consists of two nodes, two edge weights, and two time-stamps: 
\begin{enumerate}
    \item Account Pair: A node pair of two accounts can denote a directed edge $(u,v)$ with a start node $u$ and an end node $v$.
    \item Number of Transactions: A positive integer $n$, the number of transactions between the two accounts in each pair.
    \item Amount of Money Transferred: An edge weight $k$, corresponding to the total amount of money in undisclosed \$ currency, transferred from the start node to the end node from 2010 to 2020. 
    \item Start Year $Y_{1}$: The year corresponding to the first transaction for each directed edge.
    \item End Year $Y_{2}$: The year corresponding to the last transaction for each directed edge.
\end{enumerate}

\begin{figure}[ht!]
\centering
\includegraphics[width= 0.8\textwidth]{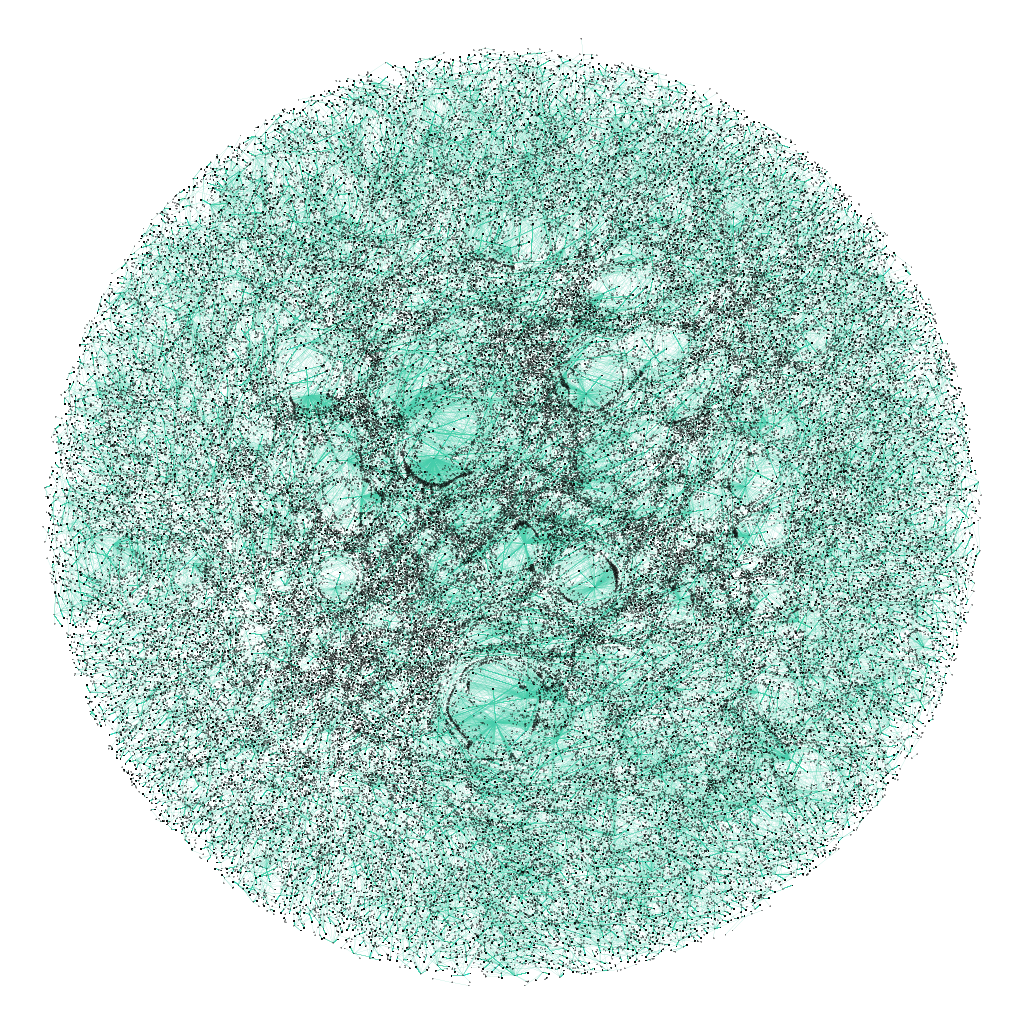}
\caption{Rabobank banking network, with edges (in green) representing transactions in the same year and under \$10,000.} \label{fig:fN}
\end{figure}

The dataset comprises 1,624,030 nodes (which represent accounts) and directed edges formed by 4,127,043 transactions. Omitting directed loops from account to themselves, there are only 3,823,167 directed edges. Saxena et al.\ studied the network’s community structure and showed that the bank's users are organized into smaller groups, with some are making more transactions between themselves and fewer transactions outside the group; see \cite{saxena}. As discussed, any transaction over \$10,000 is reported due to regulatory thresholds. In Figure~2, we see an extracted subset of 481,769 directed edges of unreported transactions; these edges represent transactions that occurred within the same year and were also under the \$10,000 threshold. Identifying directed cycles within this subset could still represent a computationally hard problem, and it would not provide much insight into the association between a group of nodes. To search within the entire set of 1.6 million accounts, we apply an algorithm that can partition the network into communities.

\section{Community and Cycle Detection}

Community structures in networks refer to observable groupings characterized by nodes that are more densely connected within their respective clusters and sparsely connected to nodes outside of them. Community detection involves finding these groups of nodes, or communities, by partitioning a network, which can help reveal underlying patterns and organization within the data. The quality of the partitions resulting from these methods is measured by the \emph{modularity} of the partition. 

Modularity is a  value $Q\in [-1/2,1)$ that measures the density of the edges inside communities as compared to the edges between communities, and it is defined as:
$$Q= \frac{1}{2m} \sum_{i,j} \left[A_{i,j} - \frac{k_{i}k_{j}}{2m}\right]\delta(c_{i},c_{j}),$$
where $A_{i,j}$ is the weight of the edge between $i$ and $j$, $k_{i}=\sum_{j}A_{i,j}$ is the sum of the weights of all edges incident to node $i$, $c_{i}$ is the community to which node $i$ is a member of, and $m = \frac{1}{2}\sum_{i,j}A_{i,j}$ is the sum of all edge weights. The $\delta$-function $\delta(u,v)$ is 1 if $u= v$, and 0 when otherwise; it will only add up terms when the two nodes $i$ and $j$ are members of the same community.

There are different strategies to increase the reliability of a partition, including modularity optimization, spectral clustering, and hierarchical clustering \cite{girvan}. Although optimizing modularity is a computationally hard problem, the goal is always to partition the network into communities that maximize the inter-community connections and minimize the intra-community connections. In 2008, Blondel et al.\ introduced an algorithm that finds high modularity partitions for large networks in a short time; see \cite{blondel}. The Louvain Algorithm is an unsupervised algorithm that iteratively optimizes the modularity score by moving nodes between communities to increase the modularity value, continuing the process until no further improvements can be made. Figure~3 shows the sixth community found from the Rabobank dataset using the Louvain algorithm. 
\begin{figure}[ht!]
\centering
\includegraphics[width= 0.8\textwidth]{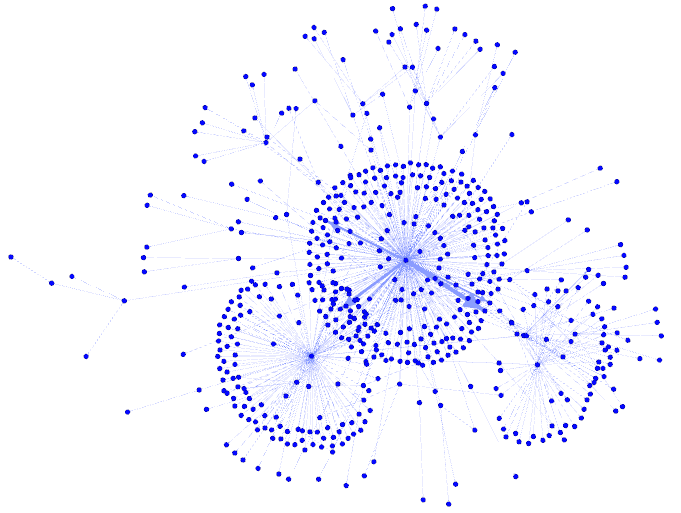}
\caption{A community in the Rabobank banking network.}\label{fig:f11}
\end{figure}

Due to the size of a banking transaction network, we will also need a directed cycle detection algorithm for the final step of our algorithm. There are multiple available options. We have used the \emph{simple\_cycles} algorithm available in Python libraries; for more information on the background of this algorithm, see Gupta et al.\ \cite{gupta}. We will now describe the combined steps that can be used to identify unusual movement of money within communities that could be flagged as potentially suspicious transactions.

\section{AML Algorithm for Suspicious Transactions} 

We aim to identify transactions within a community that could be attempting to layer and integrate illicit funds into the banking network. Additionally, we focus on transactions that are not already being reported to the regulatory agencies by excluding those over the \$10,000 threshold. Note that communities with orders fewer than three can be excluded from the analysis, as bi-directional transactions are common in banking and are not sufficient to flag an attempt at layering illicit funds. 

For a network of banking transactions $G$ with data elements of the form $\mathbf{\vec{p}}(u,v;n,k,t)$, where a pair of nodes $u,v\in V(G)$ represent two bank accounts, and $t$ represents the period during which the two accounts performed an $n$ number of transactions amounting to $k$ currency transferred. 

\smallskip

We now present our algorithm with the following steps:
\begin{enumerate}
    \item Apply the Louvain algorithm on $G$ to detect a set $\mathbf{C}_{1}$ of communities larger than two nodes.
    \item Extract a set of induced subgraphs $\mathbf{C}_2$ from each community $C_{1,i} \in \mathbf{C}_{1}$ that contains pairs of nodes with edges that have a period $t$ less than an established threshold $t_{0}$.
    \item Extract a second set of induced subgraphs $\mathbf{C}_3$ from each community $C_{2,i} \in \mathbf{C}_{2}$ that contains pairs of nodes with edges that have amounted to $k<\$10,000$.
    \item Run the \emph{simple\_cycles} algorithm within $\mathbf{C}_{3}$ to generate a set of directed cycles $\mathbf{C}_4$.
\end{enumerate}

The result provides a list of similar nodes within community structures, such as a node $u$ in $C_{4,i}\in \mathbf{C}_4$ that is contributing to the movement of an unreported amount of money across all other nodes in the cycle $C_{4,i}\in \mathbf{C}_4$, only to have an equivalent unreported amount of money returning to $u$ within a period less than $t_{0}$.

\section{Results}

The algorithm was implemented using Python on the Rabobank dataset. The initial refinement activities primarily involved formatting the data to fit the requirements of each step in the algorithm. NetworkX, a free Python package for studying networks, was utilized for various features. Specifically, the \textit{DiGraph} class from NetworkX was employed to represent the network, and the community and cycle detection functions (\textit{louvain\_communities()}/\textit{simple\_cycles()}) were used with default settings. 

As the dataset is not public, we will discuss the results of the different stages of the algorithm below. 
\begin{figure}[ht!]
\centering
\includegraphics[width=5 cm]{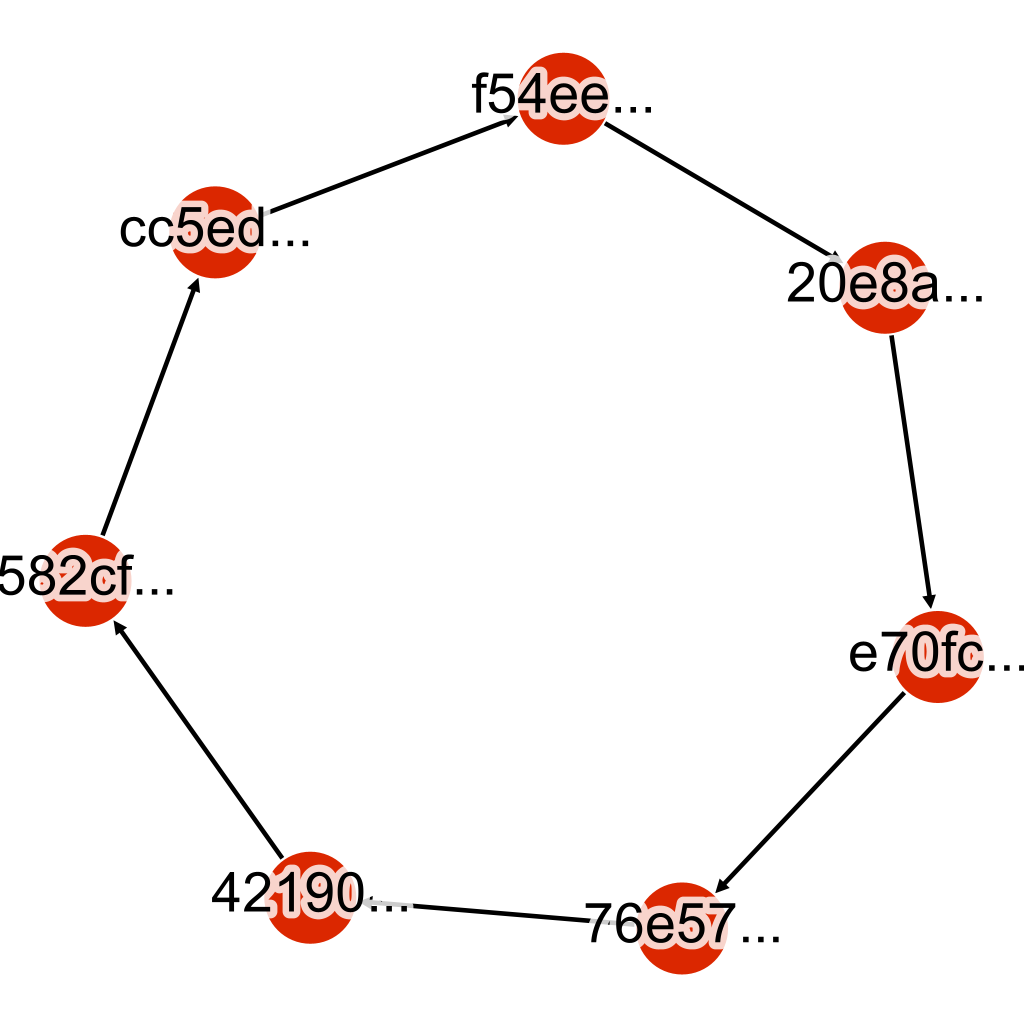}
\caption{A resulting directed cycle from the algorithm in Rabobank.}\label{fig:f12}
\end{figure}

\begin{figure}[htpb!]
\centering
\includegraphics[width=8 cm]{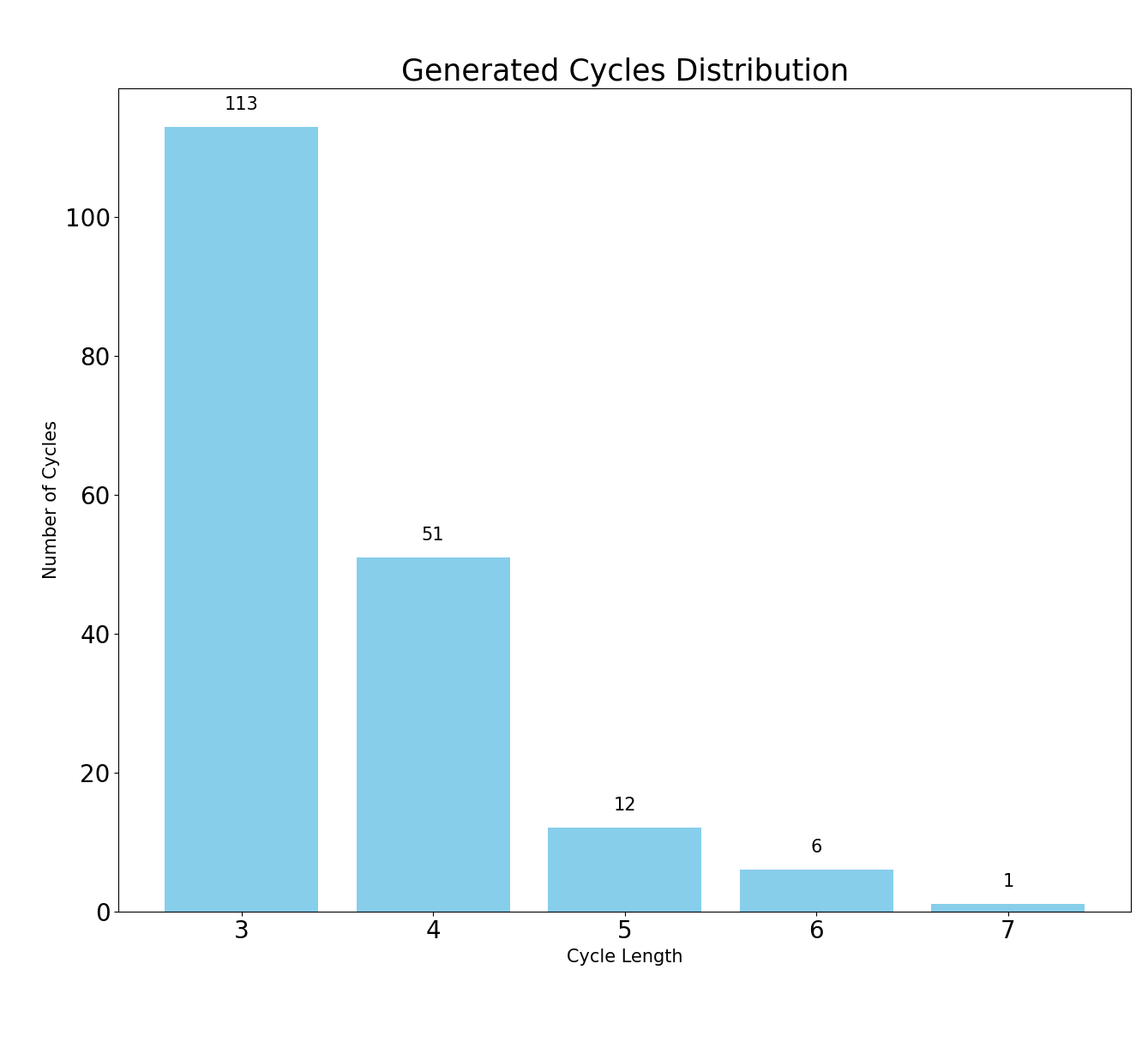}
\caption{The distribution of small directed cycles in the Rabobank banking network.}\label{chart}
\end{figure}

\begin{enumerate}
    \item The initial inputted dataset consisted of 1,624,030 nodes and 3,823,167 edges.
    \item The Louvain algorithm returned 24,292 communities of account pairs and their corresponding edges. The average size of a community was 40, and the largest community contained 5,577 accounts.
    \item Out of the total number of edges within those communities, 1,763,113 edges corresponded to transactions conducted within one year or less.
    \item Out of that subset of transactions conducted within one year or less, 1,697,761 edges corresponded to transactions that amounted to a total below the \$10,000 reporting threshold.
    \item After running the cycle detection algorithm within those communities, we identified 183 directed cycles of unreported transactions. Figure~4 shows an example of a directed cycle containing seven accounts. The distribution of the 183 generated directed cycles is represented in Figure~\ref{chart}; it should be noted that most were 3- and 4-cycles. 
\end{enumerate}

In total, 646 accounts are linked to unreported money transfers that originate in one account and, after a certain number of transactions, return a similar amount to the initial account. This behavior can be flagged for further inspection, performed through the existing suspicious transaction reports investigated by the bank's Anti-Money Laundering team to identify both the source of funds from these accounts and the purpose of the cyclic transactions. 

Even though 646 accounts represent a relatively large number to investigate, it is manageable in the current banking industry, where thousands of transactions are analyzed annually. Note that this could be a useful subset of accounts if determined to be high-risk. We can use this as an initial condition for one of the previously discussed machine learning algorithms.

\section{Discussion and Future Work}

Community detection and network analysis techniques are useful for highlighting accounts within a banking transaction network that might not be otherwise reported to regulatory authorities. Most money cycles we discovered were short, with the largest directed cycle identified as a seven-node graph. This could be due to the highly interconnected nature of a network that corresponds to the transactions within a single bank, strictly limiting the scope to the number of transaction intermediaries that use Rabobank accounts. This also aligns with Rabobank's history as a community-driven institution originally based on interconnected, locally managed cooperatives.

The fact that we did not identify thousands of money cycles is promising, as, from a practical perspective, money would rarely move in a circle within a short time. Further research could be conducted by obtaining different sets of inter-bank data, such as those available in larger quantities through international payment systems or cryptocurrency gateways. Optimization and expanded characterization are needed to adapt this model to fit a larger inter-bank environment, as there is a higher probability of identifying fraudulent activity between banks that comply with different regulators across countries than within one bank overseen by a specific regulator.

\end{document}